\documentclass{taira}
\usepackage{graphicx}
\usepackage{epstopdf, epsfig}
\usepackage{graphicx}% Include figure files
\usepackage{dcolumn}% Align table columns on decimal point
\usepackage{bm}% bold math
\usepackage{color}% bold mathfigure
\usepackage{array}
\usepackage{tabu}
\usepackage{amsmath,amssymb,stmaryrd}
\usepackage{multicol}
\usepackage{algorithm}
\usepackage[noend]{algpseudocode}
\usepackage{subfigure}
\usepackage{subfigmat,comment}
\usepackage{pbox}
\usepackage{xfrac}
\usepackage{tabularx}
\usepackage{multirow}
\usepackage[percent]{overpic}
\usepackage{float}
\usepackage{xcolor}
\usepackage{tikz}
\usetikzlibrary{shapes}

\begin{document}  

\newcommand{\sqdiamond}[1][fill=black]{\tikz [x=1.2ex,y=1.85ex,line width=.1ex,line join=round, yshift=-0.285ex] \draw  [#1]  (0,.5) -- (.5,1) -- (1,.5) -- (.5,0) -- (0,.5) -- cycle;}%
\newcommand{\MyDiamond}[1][fill=black]{\mathop{\raisebox{-0.275ex}{$\sqdiamond[#1]$}}}

\newcommand{\markerone}{\raisebox{0.0pt}{\tikz{\node[scale=0.6,regular polygon, circle, draw = {rgb,255:red,0; green,114; blue,189},line width=0.3mm, fill={rgb,255:red,0; green,114; blue,189}](){};}}}
\newcommand{\markertwo}{\raisebox{0.0pt}{\tikz{\node[scale=0.45, regular polygon, regular polygon sides=3, fill={rgb,255:red,126; green,47; blue,142}](){};}}}
\newcommand{\markerthree}{\raisebox{0.0pt}{\tikz{\node[scale=0.45,regular polygon, regular polygon sides=3,fill={rgb,255:red,119; green,172; blue,48},rotate=-90](){};}}}
\newcommand{\markerfour}{\raisebox{0.0pt}{\tikz{\node[scale=0.45,regular polygon, regular polygon sides=3,fill={rgb,255:red,77; green,190; blue,238},rotate=-180](){};}}}
\newcommand{\markerfive}{\raisebox{0.0pt}{\tikz{\node[scale=0.55,regular polygon, regular polygon sides=4, draw = {rgb,255:red,237; green,177; blue,32}, line width=0.3mm, fill={rgb,255:red,255; green,255; blue,255},rotate=0](){};}}}
\newcommand{\markersix}{\raisebox{0.0pt}{\tikz{\node[scale=0.55,regular polygon, regular polygon sides=4, draw={rgb,255:red,217; green,83; blue,25}, fill={rgb,255:red,255; green,255; blue,255},rotate=-45](){};}}}

\newtheorem{lemma}{Lemma}
\newtheorem{corollary}{Corollary}
\newcommand{\bs}{\boldsymbol}
\shorttitle{Laminar separated flows over forward-swept wings} %for header on odd pages
\shortauthor{K. Zhang \& K. Taira } %for header on even pages

\title{Laminar vortex dynamics around forward-swept wings}
% % 
\author
 {
 Kai Zhang\aff{1}
  \corresp{Email: kai.zhang3@rutgers.edu} \and
  Kunihiko Taira\aff{2}
  }

\affiliation
{
\aff{1}
Department of Mechanical and Aerospace Engineering, Rutgers University, Piscataway, 08854, NJ

\aff{2}
Department of Mechanical and Aerospace Engineering, University of California, Los Angeles, 90095, CA
}
\maketitle
%%%%%%%%%
\begin{abstract}
Forward-swept wings offer unique advantages in the aerodynamic performance of air vehicles.
However, the low-Reynolds-number characteristics of such wings have not been explored in the past.
In this work, we numerically study laminar separated flows over forward-swept wings with semi aspect ratios $sAR=0.5$ to 2 at a chord-based Reynolds number of 400.
Forward-swept wings generate wakes that are significantly different from those of backward-swept wings.
For low-aspect-ratio forward wings, the wakes remain steady due to the strong downwash effects induced by the tip vortices.
For larger aspect ratio, the downwash effects weaken over the inboard regions of the wing, allowing unsteady vortex shedding to occur.
Further larger aspect ratio allows for the formation of streamwise vortices for highly-swept wings, stabilizing the flow.
Forward-swept wings can generate enhanced lift at high angles of attack than the unswept and backward-swept wings, with the cost of high drag.
We show through force element analysis that the increased lift of forward-swept wings is attributed to the vortical structure that is maintained by the tip-vortex-induced downwash over the outboard region of wing span.
The current findings offer a detailed understanding of the sweep effects on laminar separated flows over forward-swept wings, and invite innovative designs of high-lift devices.
\end{abstract}

\section{Introduction}
\label{sec:intro}
Low-Reynolds-number aerodynamics of finite-aspect-ratio lifting surfaces have been widely studied for developing micro air vehicles and understanding biological flights.
The wakes of these small-scale wings embodies rich flow physics comprised of unsteady separation, vortex formation, and wake interactions, which are influenced by the Reynolds number, platform shape, and unsteady maneuvers \citep{mueller2003aerodynamics,buchholz2006evolution,visbal2013three,winslow2018basic,eldredge2019leading}.
The sweep angle has also been shown to play a key role in shaping the wake dynamics \citep{kuchemann1953types,harper1964review,visbal2019effect,zhang2020laminar}. Most of these studies have focused on the backward-swept wings.

Forward-swept wings, although less common in air vehicle designs, offer unique advantages that make them suitable for agile flights at high angles of attack.
In general, swept wings have the property that their aft sections stall first. 
In conventional backward-swept wings, stall occurs first near the wing tip \citep{harper1964review,black1956flow,zhang2020laminar}, resulting in a loss of aileron control.
For forward-swept wings, since stall commences at the inboard region, ailerons can still be effective at high angles of attack to offer high maneuverability.
This feature has been demonstrated by the Grumman X-29 experimental aircraft, which gave pilots excellent control response up to $45^{\circ}$ angle of attack \citep{putnam1984x}.
In addition, many aerial animals have been observed to sweep their wings forward during slow flights because the wings are required to operate at high angles of attack providing high lift to support body weight \citep{thomas2001animal}.
Many species also exhibit forward wings during high angle-of-attack perching and snatching maneuvers \citep{manchester2017variable}.

The aerodynamic benefits of forward-swept wings have attracted a number of studies on their wake characteristics.
%, especially with the development in structural strengthening techniques \citep{sherrer1981wind} that alleviate the inherent static aerodynamic divergence problem \citep{ricketts1980wind}.
\citet{breitsamter2001vortical} conducted extensive experimental investigations on the flows over forward-swept wings at $Re=4.6\times 10^5$.
They observed that the wing tip vortex and the leading edge vortex with an opposite sense of rotation dominate the flowfield. The leading-edge vortex can burst at moderate angles of attack in the outer wing region.
\citet{traub2009aerodynamic} studied these flows over a thin $65^{\circ}$ swept wing in forward and backward sweep configurations, both revealing vortex dominated flows.
Surface flow visualization showed large extents of flow separation near the wing root of the forward-swept wings, which feature significantly lower lift compared to the aft swept wings.
\citet{lee2009vortex} measured the near-field tip-vortex flow behind a forward-swept wing at $Re=1.74\times 10^5$.
Recently, \citet{setoguchi2020low} investigated the separated vortical flows from the forward-swept wings at low speed and high angle of attack conditions through Reynolds-averaged Navier-Stokes simulations.
They asserted that forward-swept wings exhibit benefits in terms of stall characteristics, since separated vortices remain on the outboard wing up to high angles of attack. 

The above studies have revealed the aerodynamic characteristics of forward-swept wings at relatively high Reynolds numbers.
However, a detailed characterization of the wake dynamics of forward-swept wings at low Reynolds number remains unexplored.
A thorough understanding of the low-$Re$ wakes of forward-swept wings is not only of fundamental importance, but also lays the foundation for design of small-scale high-lift devices.
In this work, we perform a large number of	 direct numerical simulations to study the three-dimensional separated flows over forward-swept finite-aspect-ratio wings at a Reynolds number of 400.
The present study, together with our previous work on backward-swept wings \citep{zhang2020laminar}, offers a comprehensive coverage of sweep effects on separated flows over finite-aspect-ratio wings.
In what follows, we present the computational setup and its validation in \S \ref{sec:setup}.
The results are discussed in \S \ref{sec:results}, where we provide descriptions of the wake vortical structures and the aerodynamic forces. 
We also present a force element analysis to identify key wake structures that are responsible for lift generation. 
We conclude this study by summarizing our findings in \S \ref{sec:conclusions}.

\section{Computational setup}
\label{sec:setup}

\begin{figure}
\centering
\includegraphics[width=0.7\textwidth]{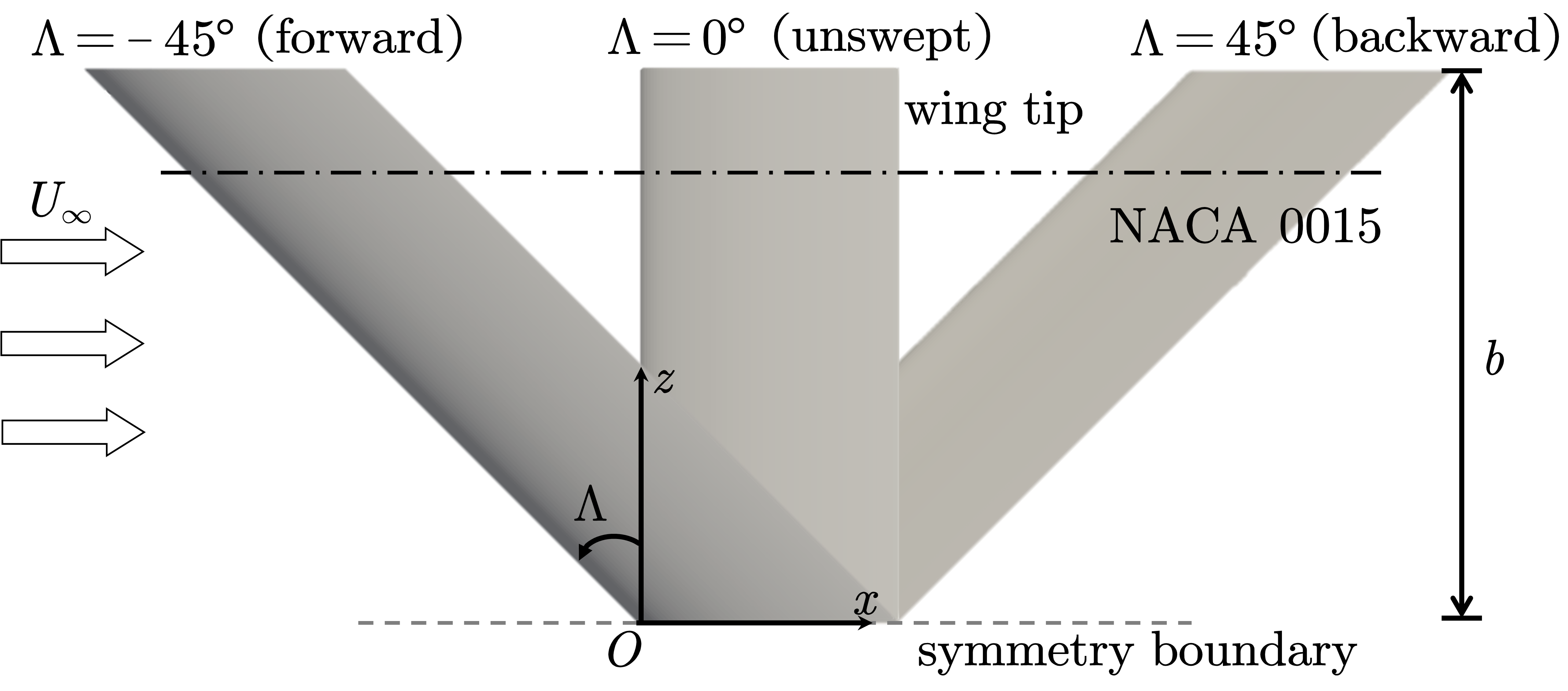}
\caption{Case setup for flows over swept wings}
\label{fig:setup}
\end{figure}

We study the incompressible flows over finite-aspect-ratio forward-swept wings with a NACA 0015 cross section. 
A schematic of the wing is shown in figure \ref{fig:setup}.
The wings are subjected to uniform flow with velocity $U_{\infty}$ in the $x$ direction. The $z$ axis aligns with the spanwise direction of the unswept wing, and the $y$ axis points in the direction of lift.
The sweep angle $\Lambda$ is defined as the angle between the $z$ axis and the leading edge.
The symmetry boundary condition is prescribed along the midspan. 
Denoting half wing span as $b$, the semi aspect ratio is defined as $sAR=b/c$, where $c$ is the chord length.
The Reynolds number, defined as $Re\equiv U_{\infty}c/\nu$ ($\nu$ is the kinematic viscosity of the fluid), is fixed at 400.
In what follows, all the spatial variables are normalized by the chord length $c$, velocity by $U_{\infty}$, and time by $c/U_{\infty}$.
The lift and drag coefficients are defined as $C_L=F_L/(\rho U_{\infty}^2bc/2)$ and $C_D=F_D/(\rho U_{\infty}^2bc/2)$, where $F_L$ and $F_D$ are the aerodynamic forces exerted on the semi-span wing in the $y$ and $x$ directions, respectively, and $\rho$ is the fluid density.

\begin{figure}
\centering
\includegraphics[width=1\textwidth]{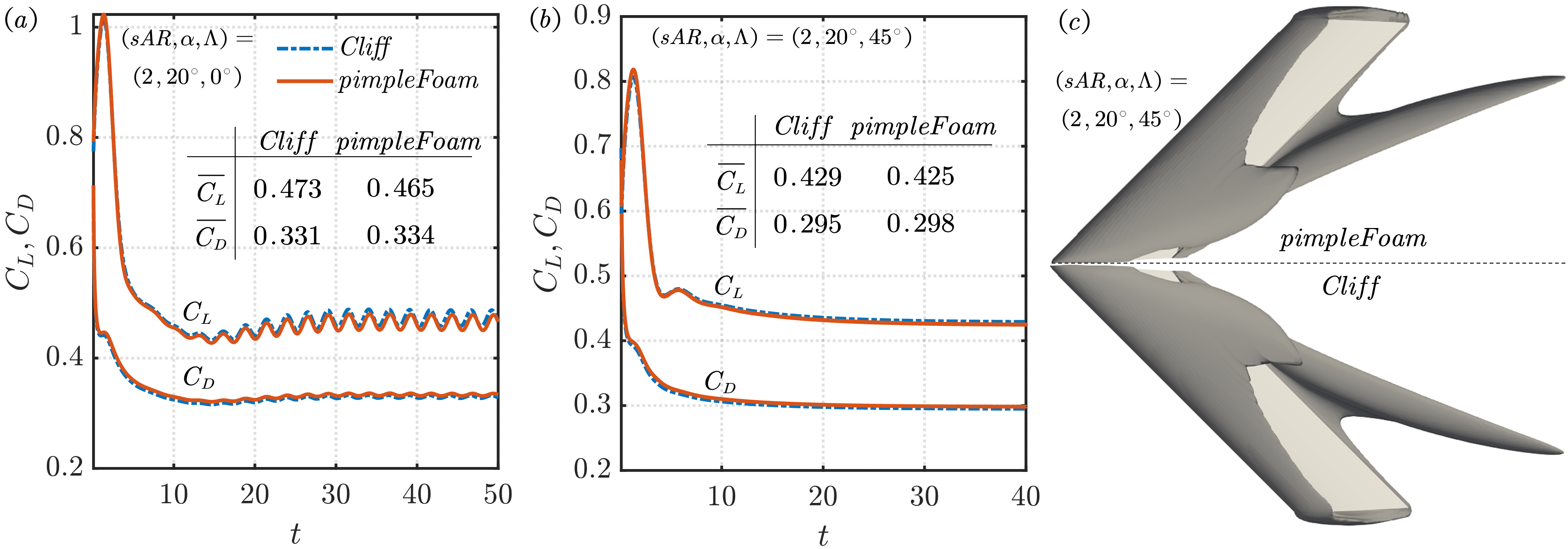}
\caption{Comparisons of force coefficients ($a,b$) and wake vortical structures ($c$) computed from \emph{Cliff} and \emph{pimpleFoam} for $(sAR,\alpha,\Lambda)=(2, 20^{\circ}, 0^{\circ})$ and $(sAR,\alpha,\Lambda)=(2, 20^{\circ}, 45^{\circ})$. The inset tables in ($a,b$) summarize the time-averaged force coefficients $\overline{C_L}$ and $\overline{C_D}$. The vortical structures in ($c$) are visualized with isosurfaces of $Qc^2/U_{\infty}^2=1$, where $Q$ is the second invariant of the velocity gradient tensor.}
\label{fig:validation}
\end{figure}

The finite-volume-based flow solver \textit{pimpleFoam} of the \textit{OpenFOAM} toolbox \citep{weller1998tensorial} is used to simulate the flows with second-order spatial and temporal accuracy.
We employed a similar grid setup from our previous work on backward-swept wings \citep{zhang2020laminar,zhang2020formation}, for which a commercial CFD code \textit{Cliff} (\textit{CharLES} package) was used \citep{ham2004energy,ham2006accurate}.
To validate the present numerical setup, we compare the force coefficients of the unswept wing $(sAR,\alpha,\Lambda)=(2,20^{\circ},0^{\circ})$ and backward-swept wing $(sAR,\alpha,\Lambda)=(2,20^{\circ},45^{\circ})$ computed from \textit{Cliff} and \textit{pimpleFoam} in figures \ref{fig:validation}($a$) and ($b$). 
The time traces of the forces obtained from the two solvers agree well with each other, and their time-averaged values differ less than 1\%. 
In addition, the vortical structures of the backward-swept wing computed from both solvers match almost perfectly as shown in figure \ref{fig:validation}($c$), validating the present computational results.

\section{Results}
\label{sec:results}

\subsection{Wake vortical structures}
\label{sec:wakes}

\begin{figure}
	\centering
	\includegraphics[width=0.9\textwidth]{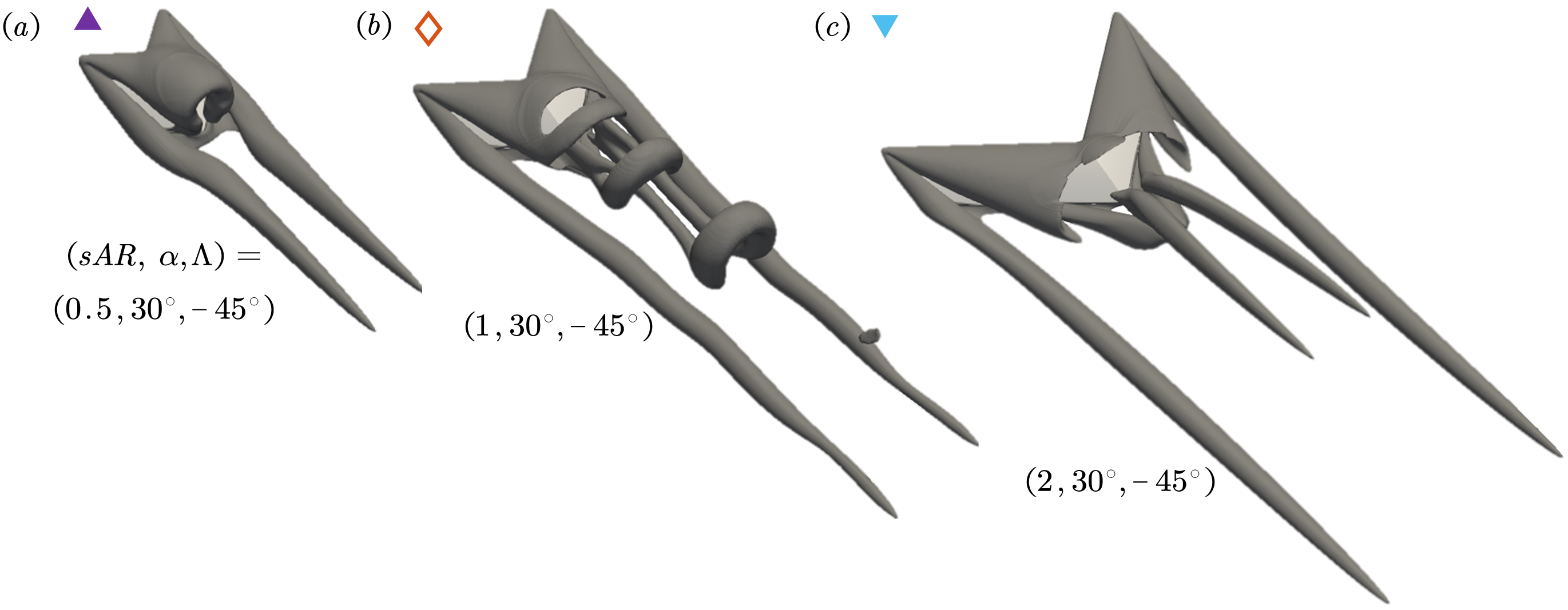}
	\caption{Representative wake vortical structures for forward-swept wings, visualized by isosurfaces of $Qc^2/U_{\infty}^2=1$.}
	\label{fig:repStruct}
\end{figure}

\begin{figure}
	\centering
	\includegraphics[width=0.75\textwidth]{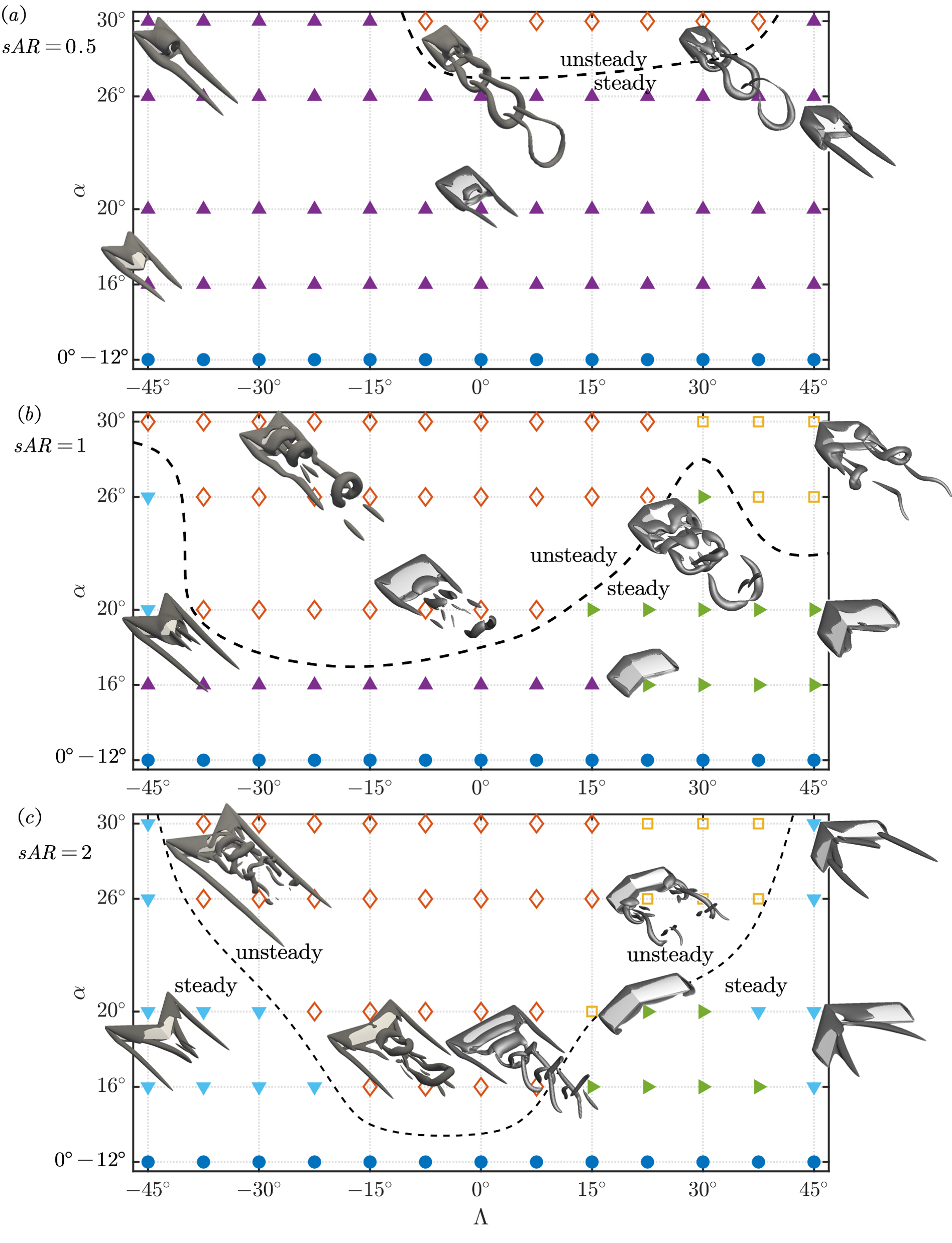}
	\caption{Classification of wake vortical structures behind swept wings for ($a$) $sAR=0.5$, ($b$) $sAR=1$ and ($c$) $sAR=2$. \protect\markertwo: steady flow due to tip effects; $\MyDiamond[draw={rgb,255:red,217; green,83; blue,25},line width=0.3mm, fill=white]$: unsteady shedding near midspan; \protect\markerthree: steady flow due to midspan effects; \protect\markerfive: unsteady shedding near wing tip; \protect\markerfour: steady flow with streamwise vortices. \protect\markerone: steady flow due to low $Re$. The dashed lines denote the approximate boundaries between steady (filled symbols) and unsteady (empty symbols) flows. The vortical structures are visualized by isosurfaces of $Qc^2/U_{\infty}^2=1$ for representative cases. }
	\label{fig:regime}
\end{figure}

The wake vortical structures behind high-incidence forward-swept wings can take three major forms, as shown in figure \ref{fig:repStruct}.
These wakes are classified as ($a$) steady flow due to tip vortices (\protect\markertwo), ($b$) unsteady vortex shedding near the midspan ($\MyDiamond[draw={rgb,255:red,217; green,83; blue,25},line width=0.3mm, fill=white]$), and $(c)$ steady flow with streamwise vortices (\protect\markerfour).
We map out these flows over the $\Lambda$-$\alpha$ space for different aspect ratios in figure \ref{fig:regime}, in which the backward-swept wing wakes \citep{zhang2020laminar} are also included.
The wakes for $\alpha \leq 12^{\circ}$ are clustered into one group as they are steady at $Re=400$ regardless of aspect ratio and sweep angle.
%The wakes for $\alpha \lesssim 16^{\circ}$ are not presented here since they are steady at $Re=400$.

The most prominent feature of the forward-swept wing wakes is the presence of the tip vortices, as shown in figure \ref{fig:repStruct}.
In comparison, the tip vortices can be suppressed for backward-swept wings.
This is due to the outboard spanwise flow induced by the backward-swept wing that counteracts the roll-up of flow around the tip \citep{zhang2020laminar}.
With increasing forward sweep angle, the tip vortices grow stronger with increasing forward sweep angle, as observed in figure \ref{fig:regime}.
The wake dynamics of low-aspect-ratio forward-swept wings are significantly influenced by the tip vortices.

At $sAR=0.5$, the tip vortices induce strong downwash effects over the entire span of the wings, acting to suppress the shedding of the leading-edge vortex sheet \citep{taira2009three,devoria2017mechanism,zhang2020laminar,zhang2020formation}.
As a result, the wakes of forward-swept wings with $sAR=0.5$ are mostly steady, as shown in figure \ref{fig:regime}($a$).
The leading-edge vortex sheet rolls up near the midspan into a dome-like shape.
Together with the pair of tip vortices, the vortical structures of these forward-swept wings resemble the shape of a joker's hat.
It is not until $\alpha=30^{\circ}$ that the unsteady vortex shedding emerge.
The forward-swept wing can achieve steady flow at smaller sweep angle than the backward-swept case, as the former is subjected stronger downwash effects from the tip vortices.

For higher-aspect-ratio forward-swept wings, the downwash induced by the tip vortex becomes weaker near the midspan, allowing the detachment of the leading-edge vortex sheet.
The flow thus features unsteady vortex shedding as shown in figure \ref{fig:repStruct}$(b)$.
This type of flow prevails for high-incidence forward-swept wings with $sAR=1$, as shown in figure \ref{fig:regime}($b$).
The stability boundary in the $\Lambda$-$\alpha$ space at $sAR=1$ is asymmetric with respect to $\Lambda=0$, with the backward-swept wings exhibiting steady flows over a larger region of the parameter space.
These steady flows of backward-swept wings are achieved due to the formation of a pair of vortical structures near the midspan on the suction side, which impose downward velocity to each other and stabilize the wake \citep{zhang2020laminar}.
Such midspan effects are localized to the inboard region of backward-swept wings.
For cases with large angle of attack and higher aspect ratio, the backward-swept wings generate unsteady vortex shedding near the tip region, which is not observed for forward-swept wings.

For $sAR=2$, the increased span allows for the formation of streamwise vortices (in addition to the tip vortices) over the inboard region, acting as a wake stabilizing mechanism for both forward- and backward-swept wings with high sweep angles.
For the forward-swept wings, two pairs of such streamwise vortices are observed for the $sAR=2$ wings.
The first pair resides in the near wake of the midspan region, and appear detached from the rest of the vortical structures as shown in figure \ref{fig:repStruct}($c$).
The other pair on trailing edge grows from the tip region and contracts towards the midspan. 
This set of streamwise vortices is similar to those found in the backward-swept cases, in which the streamwise vortices grow from midspan and trails toward the tip \citep{zhang2020laminar}. 
This difference reflects the reverse of the spanwise flow as the wings are swept from backward to forward. 
Since the formation of such streamwise vortices is universal for inclined slender bodies with inhomogeneous end boundary conditions \citep{sarpkaya1966separated,thomson1971spacing,zhang2020laminar}, the stability boundary at $sAR=2$ in the $\Lambda$-$\alpha$ space appear almost symmetric for both the forward- and backward-swept wings.

\subsection{Aerodynamic forces and moments}
\label{sec:forces}

\begin{figure}
\centering
\includegraphics[width=0.99\textwidth]{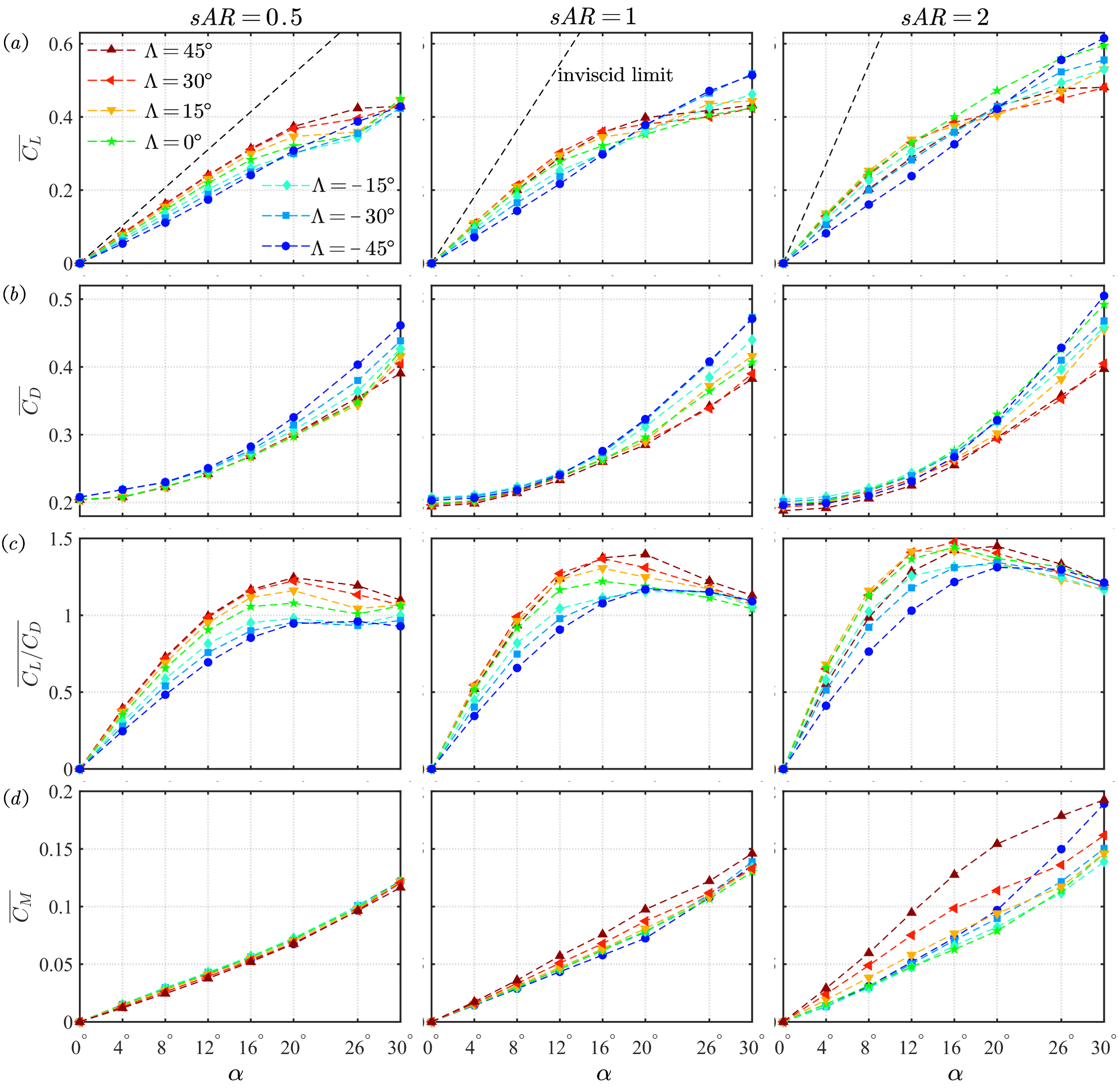}
\caption{Time-averaged aerodynamic force coefficients of forward-swept wings. $(a)$ lift coefficient $\overline{C_L}$; $(b)$ drag coefficient $\overline{C_D}$; $(c)$ lift-to-drag ratio $\overline{C_L/C_D}$ and $(d)$ pitch coefficient $\overline{C_M}$ about the quarter chord on the mid-section between tip and root planes. The black dashed lines in the $\overline{C_L}$ plots represent the inviscid limit for unswept wings. Left, middle, and right columns show data for $sAR=0.5$, 1, and 2, respectively.}
\label{fig:Forces}
\end{figure}

The time-averaged aerodynamic force coefficients $\overline{C_L}$, $\overline{C_D}$, the lift-to-drag ratio $\overline{C_L/C_D}$, and the pitch coefficient $\overline{C_M}$ of the swept wings are shown in figure \ref{fig:Forces}.
Here, the pitch moment coefficient is defined as $C_M=M/(\rho U_{\infty}^2bc^2/2)$, where $M$ is the half-span pitch moment about the quarter-chord line on the mid-section $(z=sAR/2)$ between the tip and the root planes, and positive $M$ acts to pitch the wing in the nose-up direction.
For the lift coefficients shown in figure \ref{fig:Forces}($a$), the inviscid lift limit of finite-aspect-ratio unswept wings calculated with $
C_L = 2\pi\alpha/(\sqrt{1+(1/sAR)^2}+1/sAR)$
is shown \citep{helmbold1942unverwundene}.

The lift coefficients increase almost linearly for low angles of attack ($\alpha\lesssim 12^{\circ}$).
The introduction of forward sweep has a negative impact on $\overline{C_L}$ across different aspect ratios for these low-incidence wings, while backward sweep can enhance lift for the $sAR=0.5$ and 1 wings.
On the other hand, at high angles of attack ($\alpha\gtrsim 20^{\circ}$), the forward-swept wings generally feature higher $\overline{C_L}$ than the backward-swept and unswept wings.
This phenomenon is associated with strong downwash effect provided by the tip vortices in forward-swept wings, which will be discussed in \S \ref{sec:forceElement}.

The drag coefficients shown in figure \ref{fig:Forces}$(b)$ generally exhibit an quadratic growth with angle of attack over the studied range. 
At low angles of attack, the drag coefficients remain close to each other among wings with different sweep angles.
At high $\alpha$, accompanying the high lift of forward-swept wings, $\overline{C_D}$ also increases with forward sweep angle.
As a result, the lift-to-drag ratios of the forward-swept wings are generally smaller than the backward-swept cases, as shown in figure \ref{fig:Forces}($c$).

The pitch moments for both the forward- and backward-swept wings are positive, and increase with the angle of attack monotonically, as shown in figure \ref{fig:Forces}($d$).
This suggests an inherent pitch instability with nose-up motion of the swept wings around their mid-sections.
As the aspect ratio increases, the difference in pitch moments among cases with different sweep angle enlarges.
The pitch moments of forward-swept wings are generally smaller than the backward-swept cases for wings with larger aspect ratios.
In particular, for $sAR=2$, both the forward- and backward-swept wings feature higher pitch moments than the unswept cases, indicating highly uneven force distributions along the wing span.

\begin{figure}
\centering
\includegraphics[width=0.99\textwidth]{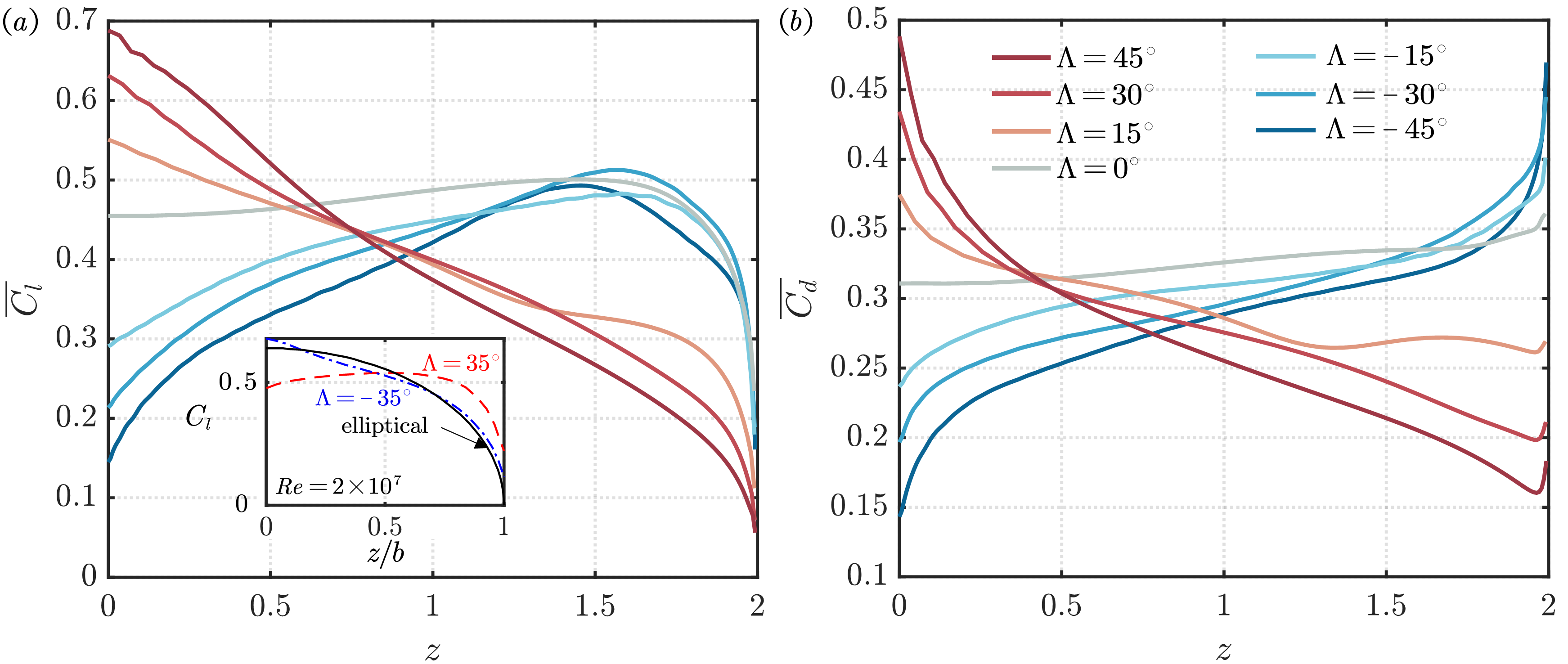}
\caption{Sectional force distributions for $(sAR,\alpha)=(2,20^{\circ})$ wings with different sweep angles. $(a)$ Sectional lift coefficients. The inset shows the lift distributions of forward- and backward-swept wings operating at $Re=2\times10^7$ \citep{gudmundsson2013general,vos2015introduction}.  $(b)$ Sectional drag distribution.}
\label{fig:sectional}
\end{figure}

The sectional lift and drag coefficients for both forward- and backward-swept wings with $(sAR,\alpha)=(2,20^{\circ})$ are shown in figure \ref{fig:sectional}.
As discussed in \citet{zhang2020laminar}, the backward-swept wings feature higher sectional lift at the inboard region, where a pair of vortical structures are formed on the suction side due to the three-dimensional midspan effects.
For the forward-swept wings, the sectional lift increases from inboard to around half chord away from the tip, and then decrease drastically towards the tip.
The high lift near the wing tip of forward-swept wings complements the low sectional lift in the inboard region, and could lead to an overall increase in lift for low-aspect-ratio wings as discussed with regards to figure \ref{fig:Forces}($a$).
The fact that the sectional lift are higher at the upwind end of the wing span (tip for forward-swept wing, and root for backward-swept wing) also explains the positive pitch moments for both wings.
We note that such lift distribution of low-$Re$ swept wings is opposite to that of high-$Re$ attached flows, in which the forward-swept wings feature higher lift inboard, and backward-swept wings features higher lift at outboard \citep{gudmundsson2013general,vos2015introduction}, as shown in the insets of figure \ref{fig:sectional}$(a)$.
Such disparity highlights the significant difference between the current vortex-dominated flows compared to the high-$Re$ attached flows.

The drag coefficients of the forward- and backward-swept wings exhibit inverse distribution along the wing span. 
In the former case, the sectional drag increases with the spanwise location and finishes with a surge towards the tip, while in the latter case, the highest sectional lift is found at the most inboard section.
The higher sectional lift and drag in the outboard region of forward-swept wings lead to increased bending moments at the root plane, calling for reinforced structural design.

\subsection{Force element analysis}
\label{sec:forceElement}
Next, let us use the force element theory \citep{chang1992potential} to identify the flow structures that are responsible for exerting lift on the wing. 
To apply this theory, we compute an auxiliary potential function $\phi_L$ (satisfying $\bs{\nabla}^2\phi_L=0$) with the boundary condition $
    -\bs{n}\cdot\bs{\nabla} \phi_L 
    = \bs{n}\cdot\bs{e_y}
    \label{equ:boundarycondition}
$
on the wing surface ($\boldsymbol{e_y}$ is the unit vector in the lift direction). By taking the inner product of the momentum equation for incompressible flow with $\boldsymbol{\nabla} \phi_L$ and integrating over the entire fluid domain $V$, the lift force can be recovered as
\begin{equation}
	F_L = \int_{V} \boldsymbol{\omega} \times \boldsymbol{u} \cdot \boldsymbol{\nabla}\phi_L \mathrm{d}V 
	+ \frac{1}{\Rey}\int_{\partial V}\boldsymbol{\omega} \times \boldsymbol{n} \cdot (\boldsymbol{\nabla}\phi_L + \boldsymbol{e_y}) \mathrm{d}S.
	\label{equ:liftElement}
\end{equation}
The integrands in the first and second terms on the right hand side of this equation are the volume and the surface lift elements, respectively. 
At $Re=400$, the volume force elements contribute more significantly to the total force than the surface force elements.

\begin{figure}
\centering
\includegraphics[width=0.99\textwidth]{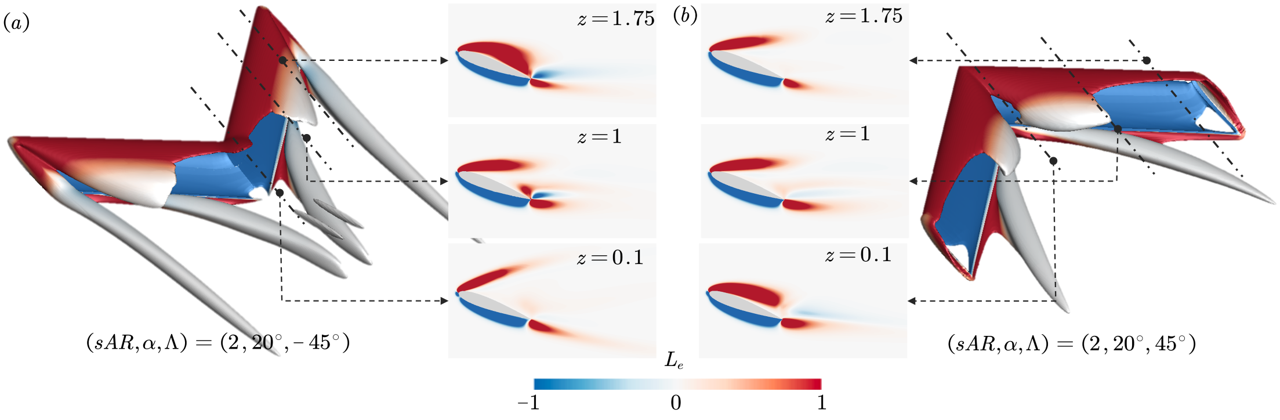}
\caption{Lift elements for ($a$) forward-swept wing $(sAR,\alpha,\Lambda)=(2,20^{\circ},-45^{\circ})$ and ($b$) backward-swept wing $(sAR,\alpha,\Lambda)=(2,20^{\circ},45^{\circ})$. The lift elements from -1 (blue) to 1 (red) are plotted on the isosurfaces of $Qc^2/U_{\infty}^2=1$. The sectional slices shows colormaps of $L_e\in[-1,1]$ at different spanwise locations.}
\label{fig:liftElement}
\end{figure}

We present the lift elements for forward-swept $(sAR,\alpha,\Lambda)=(2,20^{\circ},-45^{\circ})$ and backward-swept wings $(sAR,\alpha,\Lambda)=(2,20^{\circ},45^{\circ})$ on the isosurfaces of $Qc^2/U_{\infty}^2=1$ in figure \ref{fig:liftElement}. 
In general, the vortex sheets emanating from both the leading and trailing edges are associated with positive lift, while the flow near the pressure side of the wing contribute to negative lift \citep{menon2021significance}.
For the forward-swept wing shown in figure \ref{fig:liftElement}($a$), the positive lift elements on the suction side of the wing appear thicker and closer to the wing surface towards the wing tip than those in the inboard regions. 
This is due to the strong tip-vortex-induced downwash effects that confines the vorticity in the vicinity of the suction side of the wing, as discussed in \S \ref{sec:wakes}.
On the other hand, the backward-swept wing enhances lift by the formation of a pair of vortical structures near the midspan, where the concentrated lift elements are attached to the wing surface \citep{zhang2020laminar}.
Despite these different mechanisms in maintaining the steady vortical structures on the suction sides of the wings, both forward- and backward-swept wings are able to harness separated flows to generate additional vortical lift at high incidence, thus enhancing their aerodynamic performance.

\section{Conclusions}
\label{sec:conclusions}
We have performed three-dimensional direct numerical simulations to study the effects of forward sweep on the wake dynamics and aerodynamic characteristics of finite-aspect-ratio wings at a chord-based Reynolds number of 400, covering aspect ratios of $sAR=0.5$ to $2$, angles of attack $\alpha=0^{\circ}$ to $30^{\circ}$, and sweep angle $\Lambda=0^{\circ}$ to $-45^{\circ}$. 
The flows over forward-swept wings generally feature a pair of counter-rotating tip vortices, which shapes the wake dynamics significantly.
For low-aspect-ratio forward-swept wings, the wakes are strongly influenced by the tip-vortex-induced downwash effects and remains steady. 
With increasing aspect ratio, the downwash effects weaken over the inboard regions of the wing, resulting in unsteady vortex shedding.
For wings with higher forward sweep angles, streamwise vortices develop in the wake, again stabilizeing the flow.
While the introduction of forward sweep can have a negative impact on the lift at low $\alpha$, the forward-swept wings can experience enhanced lift from the separated flows, resulting in higher lift than the backward swept wings.
This comes at the cost of high drag, which results in lower lift-to-drag ratios for forward swept wings.
Through a force element analysis, we show that the high lift of forward swept wings is mainly contributed by the steady vortical structure near the tip region, which is formed under the downwash induced by the tip vortices.
Together with our previous studies on backward-swept wings, the present results provide a comprehensive understanding of the sweep effects on the laminar separated flows over finite-aspect-ratio wings and offer coverage of less-explored area of low-$Re$ aerodynamic database.

%\section*{Declaration of interest}
%The authors report no conflict of interest.

\section*{Acknowledgement}
KZ is grateful for the Office of Advanced Research Computing (OARC) at Rutgers, The State University of New Jersey for providing access to the Amarel cluster.
KT acknowledges support from the US Air Force Office of Scientific Research (FA9550-17-1-0222 and FA9550-22-1-0174).

\bibliography{jfm-reference}
\bibliographystyle{jfm}
\end{document}